# DDNFS: a Distributed Digital Notary File System

Alexander Zangerl

Bond University, School of Information Technology

**Abstract**

*Safeguarding online communications using public key cryptography is a well-established practice today, but with the increasing reliance on "faceless", solely online entities one of the core aspects of public key cryptography is becoming a substantial problem in practice: Who can we trust to introduce us to and vouch for some online party whose public key we see for the first time? Most existing certification models lack flexibility and have come under attack repeatedly in recent years[1, 2], and finding practical improvements has a high priority.*

*We propose that the real-world concept of a notary or certifying witness can be adapted to today's online environment quite easily, and that such a system when combined with peer-to-peer technologies for defense in depth is a viable alternative to monolithic trust infrastructures.*

*Instead of trusting assurances from a single party, integrity certifications (and data replication) can be provided among a group of independent parties in a peer-to-peer fashion. As the likelihood of all such assurance providers being subverted at the very same time is very much less than that of a single party, overall robustness is improved.*

*This paper presents the design and the implementation of our prototype online notary system where independent computer notaries provide integrity certification and highly-available replicated storage, and discusses how this online notary system handles some common threat patterns.*

## 1 Introduction

The need for notaries, Justices of the Peace and similar certifying witnesses in the real world is wellestablished and understood, whereas in the online universe we experience an ever-increasing number of situations where similar online services are completely missing or substantially lacking[3].

This paper presents our contribution towards mitigating this problem: We propose that an online digital analog of the real-world notary system can provide a valuable mechanism for making the online universe a safer and more trustable environment.

### 1.1 Real-World Certifications

The common existing real-world certification means aren't exactly perfect - one has to worry about weak signatures, rubber stamps that are easy to forge etc. - and far from uniformly available across the globe or governed by similar rules.

Nevertheless we use them - for practicality's sake, because humans are relatively good at being suspicious, and because it's always possible to request further, possibly higher-quality certifications when there is any doubt.

But different from the real world, where humans get their cues regarding trustworthiness from a variety of sources, in the online universe there is no such direct interaction between parties, and fuzzy degrees of mistrust are hard to come by: computers do precisely what they're programmed to, without question and regardless of where the instructions come from - and making computers suitably suspicious and hard to trick is not easy.

## 1.2 Current Online Certification Schemes

Providing any sort of integrity assurance and provenance of online materials involves trusting somebody's word, e.g. the original source's - but with which we never have any direct contact.

One practical compromise for this faceless nature of online communications that is commonly used used today is to offload trust and verification aspects to third parties and rely on their diligence. In the SSL security model (used for secure web access amongst other services), trust is founded on absolutely trusting commercial Certification Authorities (or CAs).

But recently the number of incidents concerning CAs has risen sharply[4, 5], and the trustworthiness of many CAs has to be called into question[2, 1]. From a practical and global perspective the monolithic and absolute SSL trust model is crumbling badly.

Alternative trust environments like the PGP Web of Trust[6] exist and thrive, but in a less commercial environment than the WWW and requiring substantially higher degrees of effort by the users. Recently a few projects like Monkeysphere[7] and networknotary.org (now known as the "Perspectives Project"[8]) have appeared, which try to bridge the gap between current practices and higher degrees of security in a meaningful fashion.

## 1.3 Defense in Depth

So what can be learned from the real-world notary systems to possibly improve the online situation? We believe that one of the most important aspects that need improvement is the possibility of using multiple certifying parties, which are as independent as possible and where each needs to be trusted to a limited degree only. This is very much along the lines of the security tenet that all resources need to be defended in depth, by multiple, often overlapping security mechanisms.

Applied to certifying data provenance this means that a certification's strength can rely in part on the fact that it is very unlikely that all the involved certifiers are subverted and colluding to falsify information (cf. the Byzantine Generals problem[9, 10]).

This multiple layer approach makes especially good sense in the online world, where services can be available all the time and on demand and where incorporating extra certifying agencies is not overly onerous.

To demonstrate this idea we have developed a mechanism for tamper-resistant online data storage with strong integrity assurance that is provided amongst any number of slightly cooperating but otherwise independent peers. We have also implemented a testbed application with a very convenient user interface to ease deployment and usability: the ddnfs application presents itself to users as an overlay file system, which handles both the retrieval of data and certifications from other notaries on demand as well as the submission of newly created material for certification, all in a manner that is transparent to the user.

The remainder of the paper is organized as follows: Section 2 describes relevant related work, followed by Section 3 which outlines the design and architecture of ddnfs. In Section 4 we discuss how ddnfs deals with security threats and Section 5 briefly describes the implementation of our service prototype.

## 2 Related Work

At the core of the ddnfs environment lies the problem of replicating and certifying data in an efficient and safe manner amongst untrusted and independent peers. One of the most interesting systems for handling such relaxed-consistency replicated data is called Bayou which is part of Alan Demers' works at Xerox PARC. While Bayou does not take malicious failures into account and is by itself therefore insufficient for Byzantine environments (which ddnfs aims for), the architectural paper[11] outlining the goals and design decisions underlying the Bayou system was very influential for ddnfs.

Bayou strives to support data sharing among mobile clients, with the explicit assumption that clients may experience extended and involuntary loss of communications. To provide high availability despite such outages, the Bayou system employs weakly consistent replication of data. Bayou tries to minimize assumptions about the available communication infrastructure by using peer-to-peer "anti-entropy" for propagating updates, which involves all replicas regularly doing a pair-wise synchronization with a randomly chosen other party. This approach relies on almost no infrastructure whatsoever, an aspect that has become a recurring theme with the recent commonality of peer-to-peer systems[12].

To maximize a user's ability to read and write data, Bayou allows clients to write to any replica; this of course creates the possibility for conflicting operations being inserted at different peers which have to be resolved somehow. Acknowledging and embracing the floating nature of data updates, Bayou deals with such update conflicts by allowing application-specific automatic merge procedures. Among other important ideas Bayou also introduced the notion of tentative versus committed writes: committed writes are stable, with no tentative writes being ordered before them. This can be achieved by distributed consensus among the parties (which is where the Byzantine Generals problem awaits the implementor), or by restricting operations slightly as in Bayou's case: here each database has a "primary" server that is responsible for committing writes. Bayou's authors decided on this simpler operation in order to get away with minimal communication guarantees. In Bayou (as in ddnfs), clients can read tentative data if they so desire; this is allowed to keep Bayou simple yet flexible and to allow applications to select their own appropriate point in the trade-off between consistency and availability.

In a subsequent paper[13], Petersen et al. expanded on the details and resulting characteristics of the Bayou system and also present one of the few practical application examples: they implemented BXMH, a version of the EXMH[14] mail user agent which manages emails stored using Bayou rather than a local file system. Experimental results show practically sustainable performance with a very small code base (under 30k lines of POSIX C). Bayou comes pretty close to a minimal yet practical system for replicated shared data - except where stringent security is concerned as it covers only untrusted network infrastructure but not peers acting maliciously.

Looking at rumour mongering as a potential mechanism for epidemic distribution of data, the paper of Karp et al. [15] comes to mind as it provides an excellent coverage of the rumour mongering problem domain. That work discusses the very generic setting of spreading rumours in a distributed environment where communication is randomised and parties are independent. The goal is distributing rumours from one originator to all parties using as few synchronous rounds and messages as possible, and some exchange protocols are suggested and analysed, allowing for what they called "adversarial node failures" and which does not encompass Byzantine failures but rather a set of faulty peers which fail to exchange information in some of the communication rounds. This falls short of Byzantine safety as exchanging misleading or false information is not dealt with.

Nevertheless the insights provided are substantial, among which is a proof that rumour mongering algorithms inherently involve a trade-off between optimality in time and communication cost. Also, the combination of pushing and pulling exchanges is studied, with a practical algorithm being suggested. Push transmissions (sending "hot" or interesting rumours from the calling party to the callee) have the useful property of increasing the set of informed parties exponentially over time until about half of the group is involved; thereafter, the residual set shrinks at a constant factor. On the other hand, pull transmissions (asking the callee for news), are advantageous if and when most parties are involved and if almost every player places a call in every round, as then the residual set shrinks quadratically. This dichotomy suggests combining push and pull mechanisms, or at least switching between them. The decision of when to switch algorithms (and in general when to terminate the rumour mongering altogether) is the crucial aspect, controlling the robustness of the overall scheme. This is complicated when parties are independent and only have access to local approximations of the global situation at any particular time.

[15] presents some discussion of the simplest possible push-pull combination, namely sending rumours in both directions in every round; in this case rumours include an age counter to be transmitted with the data, which is used to estimate the set of informed players and controls the termination of the distribution. There are some issues with this round counter: first of all, an adversary could lie and adversely affect the distribution. Second, having a fixed maximum age is not robust – but setting this limit higher than strictly needed wastes communications. Karp suggests a scheme which transmits not just an age counter but also some state between parties, where each party uses the state messages received over time to control its switching between push-pull to pull-only and subsequently to silence, i.e. termination of the rumour spreading.

The notion of rumours carrying some meta-information about the rumour in question is also used in [12], and we have adopted a more failsafe variant of this meta-information as in our environment information passed by some party can not simply be taken at face value. ddnfs therefore uses verifiable rumours as detailed in Section 3.3.

The above-mentioned influential work on epidemic distribution by Datta, Hauswirth and Aberer[12] aims to extend epidemic algorithms towards a peer-to-peer paradigm by suggesting algorithms that tolerate highly unreliable, ephemeral peers. The paper studies the problem of updates in decentralised and self-organising peer-to-peer systems, where progress is hampered by both low probability of peers being reachable as well as the lack of global knowledge to base distribution decisions on. Their overall goal was a distributed storage scheme, with the lack of global knowledge and the need to self-organise locally making up for most of the challenge.

In their paper, Datta et al. present a hybrid push-pull strategy which is a drastically simplified variant of the protocol by Karp[15]. Their problem domain and solution approach centers on the primary assumptions that all peers are autonomous and independent without global knowledge; that probabilistic consistency is sufficient and that the peer population is large with high connectivity factors - but possibly low individual probability of a peer being online, and finally that expected failures are of the benign "failstop" kind.

Apart from aiming to protect from Byzantine failures, a number of ddnfs's design decisions do parallel the work by Datta et al.. They realise that rumour mongering would have quite limited efficiency in an environment where most systems are expected to be unaccessible at times, therefore a push-pull mechanism was adopted: the main distribution model is a push model, but peers who miss an update can use the subsequent pull phase to catch up with developments. We have adopted this approach for similar reasons.

In addition, their algorithm uses an extra mechanism, besides the usual feedback to control the spreading of information: in their system, every message carries a partial history of peers who have been offered this datum previously. This information is used to direct updates more precisely but also as a local metric of the "reach" of the distribution, which in turn controls probabilistic parameters of the rumour spreading; overall it reduces duplicate messages greatly. This feed-forward mechanism allows peers to collect a rough approximation of the global state over time, which was a novel contribution for epidemic algorithms (as well as their analytical modelling for rumour mongering).

In their algorithm, the push-phase is the predominant means of distributing information: as explained in [15], exponential growth of informed parties allows a rapid spread among online parties. The push phase is meant to allow resurrecting parties who missed information during their downtime to resynchronize at will; after the push phase, most nodes will have the information so pulling in by the remaining nodes is vastly more efficient than prolonging the first phase. The pull decision is made locally, governed by a variety of heuristic means (e.g. realization of having been off-line, not receiving any updates for a prolonged period or receiving a pull request for a dormant update).

The push algorithm exchanges messages, round counters and a list of peers that the update had been offered to by the sending party, but that does not necessarily mean the message has been received by said peers. Duplicate updates (i.e. all aspects of the message including the history list) are ignored; otherwise the receiving peer makes a probabilistic decision on further propagation of this message and if positive, increments the round counter, selects a number of candidate peers at random from the uncontacted remainder of the population, and sends updates to them. The number of push rounds controls how closely the probability of reaching the whole online population will approach unity.

Our problem scenario is subtly different, as information received by peers requires closer scrutiny. In our design the feed-forward idea was extended, using verifiable history information as described in the following section. Another important difference is that in our environment, peer identities need to be more permanent because there is an association between trust and a peer's identity.

## 3  Design

Different from other certification systems (where documents to be signed are handled out of band, e.g. PGP's web of trust) or secret sharing infrastructures[16] (where integrity of materials is preserved absolutely at the cost of limited redundancy), we decided to combine both document exchange and certification into one integral infrastructure. This stems from our original motivation for this project, which was to extend file integrity checking systems beyond a merely local scope, and which requires a mechanism for replicated, tamper-resistant storage of file fingerprint data. We aimed to minimize trust and infrastructure requirements, and hence developed a system where replicated data acquires certifications "en passant", during the exchange process and by every involved peer - which closely resembles a multi-party, multi-stage notary environment. The resulting exchange protocol also serves to certify peers' actions which makes the ddnfs infrastructure to some extent self-regulating.

We start our design overview by outlining the components of our system:

*Peers:* are independent computer systems which run the ddnfs application, and exchange and certify documents to provide tamper-resistant storage to each other.

*Network:* Any untrusted Internet-like network that the peers have access to. Permanent access is not required.

*Documents:* are the stored objects that ddnfs replicates. Documents are fixed-size and immutable once generated, and identified using a path name and a version. Newer versions of documents can supersede earlier versions under conditions that are specified in a document policy.

*Policy:* defines rules for integrity protection, required degree and numbers of certifications and authorship restrictions for particular documents.

*Administrators:* Human parties who can control the ddnfs system remotely by joining as temporary peers.

*Document Distribution Protocol:* A hybrid push/pull epidemic flooding algorithm that is used to access and distribute documents among peers. All communications between peers are encrypted.

*Signatures:* Hierarchical, over-lapping signatures on documents are exchanged and scrutinized by peers to prove a peer's acceptance of the document (content and history). These signatures provide integrity confirmation and also control the distribution process among peers.

## 3.1 Peers

Parties in the ddnfs environment are not under any kind of central control, can be widely spread geographically and communicate over unreliable public networks. Peers are independent and autonomous, and are not trusted: any peer decides for itself how far data has to be integrity-certified until the peer is satisfied and willing to use it.

This decision and others related to the document distribution operation are made locally based on independently verifiable information in the form of cryptographic signatures on data.

Correct peers are expected to cooperate and contribute somewhat altruistically towards the overall goal by signing all data they distribute, and using these cryptographic signatures allows the overall system to degrade gracefully and safely in the face of corruption of any limited number of involved parties. Subverted peers can act maliciously without this having any negative effect on the safety and integrity of other peers' data (the worst case scenario results in temporary unavailability of data or recent certifications).

Only a peer's behaviour as observed by others can be used to prove its compliance and thus trustworthiness. Correct operation is measured against compliance with what we call the Good Citizen Criteria:

- Every peer handling a document must vouch for it by cryptographically signing the document and its distribution history to provide proof of its integrity and provenance. Only signed material with verified signatures may be exchanged.
- Data exchanges are done redundantly to sufficiently many peers in parallel so that no (small) number of corrupt peers could block updates from reaching other correct peers.

## 3.2 Documents

ddnfs operates on documents, which have a body of opaque data and a path name a version associated with them. Name and version together identify a document and its content, which cannot change after initial creation. ddnfs presents these documents to users in the form of a strict versioning file system where every document change leads to a new version being created.

Also associated with each document is a set of cryptographic signatures which spans the document's body, name and version together with a certain subset of signatures present at the time of signing, which is described further in section 3.3.

Except for policy restrictions, any peer can create and offer documents to the group for replication. For obvious reasons modification of an existing document is not supported, but obviously some means for phasing out data over time is required and furthermore some crucial data may very well require human supervisory involvement.

These considerations suggest three distinct states for a document: pending, active and superseded. A document version is considered pending after its introduction to the peers until it has been distributed and certified sufficiently to fulfil the policy requirements. Pending documents are exchanged between peers in an aggressive, push fashion and acquire certifications during these exchanges. Once sufficiently signed and distributed, a document is considered active and aggressive push distribution ceases. If a newer document version becomes active, then all older versions are considered superseded and peers may at their discretion stop storing and offering these. The requirements for document activeness are expressed in the policy which is discussed in section 3.4. Figure 1 gives an overview of the possible state transitions.

## 3.3 Document Certification

Integrity verification and certification is performed by all peers as part of the document replication procedures; additionally, human administrators can provide further assurances for specific crucial data items.

This certification involves creating cryptographic signatures of the documents in question, to be performed by each peer independently. Because of the cost of creating and verifying signatures, the number of signatures should to be limited, ideally to one per peer. This approach allows a simple setup for representation and coverage of signatures, which combines the benefits of signing only once with a limited but practical degree of tamper-resistance for the transfer of signatures.

In ddnfs this is achieved with hierarchical signatures, which span a subset of signatures as present at signing. Every peer must sign once upon first reception of a document, and create a signature that covers the document content, name, version and size, and also the first observed distribution path, as given by the signatures "up-tree" of the receiving peer. Every peer's signature thus covers all the parents' signatures up to and including the originator signature.

Figure 2 shows the accumulation of signatures at a peer F over time: after the second round of distribution steps, F has received the datum from A and wrapped it in its signature. Step 3 sees D offering its knowledge, consisting of signatures by the originator, C and D. This new information is verified and then merged into an overlapping tree, without F signing again.

Subsequent document exchanges are not embedded in new signatures, therefore the design gains the benefits of a signature block that is bounded in size by the group, as well as a caching opportunity for signature verification. Furthermore the topology of this signature tree is used for heuristic fine-tuning of the propagation strategy: peers who have not signed a document yet can be prioritized for distribution.

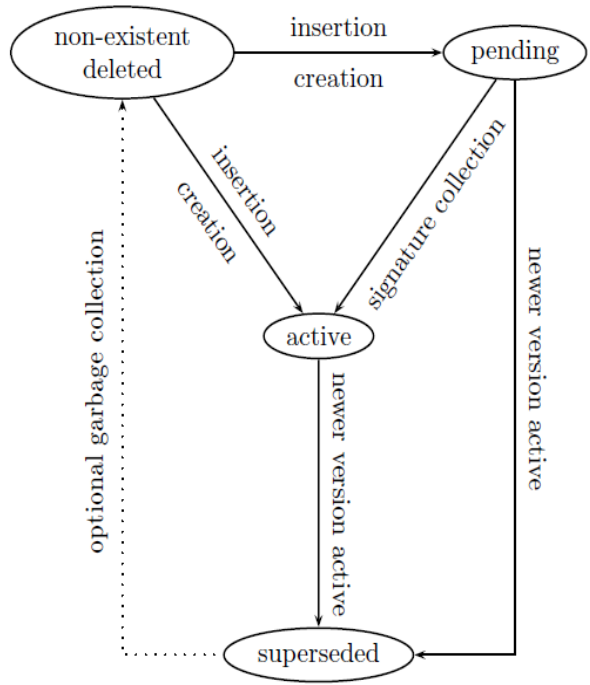

Figure 1: Document Lifecycle

The overlapping nature of signatures makes tampering with the replication process (for example by removing signatures enroute) substantially harder and can be considered good value as it incurs no extra cost: every peer must generate at least one signature anyway, and the composition of the material to be signed is straight-forward and does not involve any expensive or cryptographic operations.

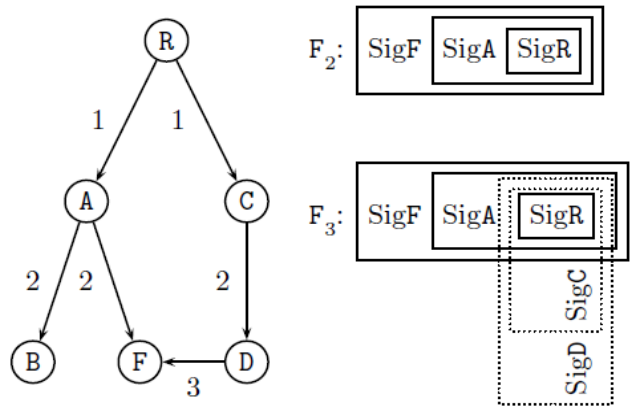

Figure 2: Accumulation of Signature Tree at Peer F

### 3.4 Control Mechanisms

The ddnfs system is configured using one special document of global scope called the peerlist. It contains the group meta data, namely the peers' public keys and network identifiers and a policy component.

The purpose of the policy is to define any restrictions on authoring of particular documents, and to set the conditions under which a document is regarded as active, i.e. when a document in a particular name space is sufficiently certified regarding its integrity.

The policy rules are expressed in a simple language which specifies for document path names a set of authorized creators of the initial signature (and thus the allowed authors) and the requirements for document activeness. The activeness criterion is comprised of an arbitrarily complex boolean expression of sets of peers who have to receive and sign the document before activeness is achieved and the distribution efforts can be switched from active push-based operation to passive.

While ddnfs is meant to be as self-contained and automatic as possible, it shouldn't be left to operate without human supervision or control. We therefore incorporated the notion of human administrators as a core component for the system, but without relinquishing flexibility: certification by administrators is an optional requirement that the policy may stipulate for a document.

Human administrators are regarded as temporary peers: like ordinary peers they do have private/public key pairs but are not involved in the push-based document distribution. Administration duties can be performed from any computer system that carries the ddnfs software and which is capable of contacting at least one member of the peer group.

In the ddnfs environment, peer group membership is not ad-hoc as in ephemeral P2P systems, because it is necessary to rely on permanence of identity and group membership in order to be able to trust peers document certifications. Group membership in our system is therefore strictly controlled: administrators decide on membership changes by creating suitable peerlist documents and submitting them to the peer group. That way membership information, administrator privileges and policy constraints are all communicated simply and efficiently, in-band using the same document distribution an and assurance provisions that are available for ordinary documents.

### 3.5 Document Distribution and Retrieval

The conceptual basis of our document distribution protocol is a hybrid push/pull epidemic algorithm, similar to rumour mongering as pioneered by Bayou[11] but with the crucial difference of transporting verifiable rumours. Any peer can create a new document but must claim responsibility for it by creating the first signature. It then offers the document's signature block as an abstract to other peers. The recipients of the offer then decide how to proceed, based on policy and local criteria: every offer received triggers a number of new offers sent out by the recipients, until the document has reached active status and offer-based push replication is terminated. As documents are signed only at most once by a given peer, the size of the overall signature block is of finite (and practically very modest) size.

Our system is fault tolerant against Byzantine Failures[9]. This requires high redundancy in the replication process and can only be achieved with a significant number of communication messages, and like many others[12] we settled for eventual consistency as a practical compromise.

The network communication environment for ddnfs involves two protocol levels: a low-level protocol provides encrypted and authenticated bidirectional unicast byte-streams connecting exactly two peers. The network connecting the peers is assumed to be a public network like the Internet, untrusted and without throughput or reliability guarantees. Communication between peers is safeguarded from tampering or eavesdropping by using symmetric cryptography on the lower protocol level.

The high-level protocol implements the necessary document-oriented operations. It is symmetric and uses tagged messages (similar to the IMAP protocol[17]) to allow interleaving of multiple requests. Requests and responses are associated by their shared tag. This protocol is very simple and comprised of just five message types: IHAVE, GET and GETANSWER and HEAD and HEADANSWER.

GET and HEAD offer retrieval commands for both signature blocks and full documents. Their semantics are simple: when a GET is received, the receiver is expected to retrieve the requested document from its local store and respond with a GETANSWER that contains the document and its signature block. Requests for non-existent documents are answered with an error indication. HEAD is similar, but only retrieves the signature block. The addressing mechanism for HEAD and GET requests includes document path name and version, and wildcards are supported for paths and to a smaller extent for versions ("any version of this document" and "the currently active version").

IHAVE is the core protocol component for performing the document distribution, and its semantics were inspired heavily by the NNTP protocol[18] (where the name was borrowed from, too). IHAVE is asynchronous, reply-less and has two main purposes:

- In the initial phase of document distribution it serves to notify peers of the existence of a new document (version) at the sender.

- During the subsequent consolidation phase IHAVEs are used to present a document's signature block to peers to spread new signatures.

On reception of an IHAVE message a peer is expected to verify and consolidate the signature state for the document in question. If the recipient knows more signatures for the document then it's expected to reflexively send an IHAVE with that larger signature set back to the sender. If the document in question is locally unknown, then the recipient is expected to follow up with a GET request. When that succeeds the recipient must first verify all signatures on the document, then add its own as outlined in section 3.3 and finally send out IHAVEs of its own to notify others of this newly created signature. All peers thus contribute to the document's certification as depicted in Figure 3.

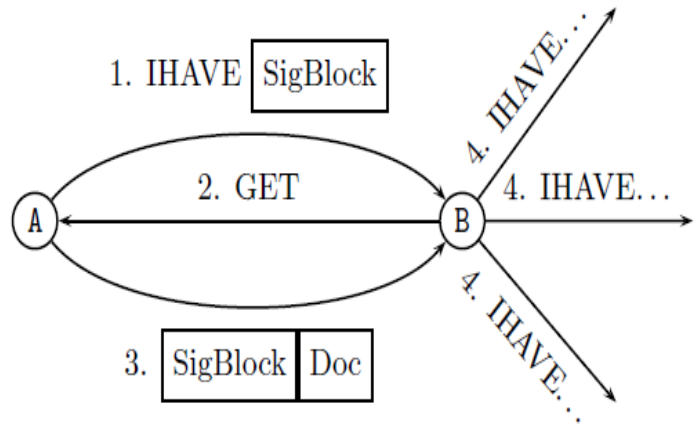

Figure 3: Initial Document Injection Phase

When the signature set available to peers is sufficient to meet policy rules for activeness (or when no uninvolved peers are left), then this distribution process stops gradually (as peers do not generally gain the full set of signature knowledge at the same time). In prior work[19, 20, 21] we have shown that relatively low degrees of offer parallelism combined with adaptive heuristics performed by peers to select who to send offers to result in this protocol performing quite favorably wrt. computational and communicational costs.

Also of interest is the fact that peers who weren't reachable during a document distribution phase can quickly and easily "catch up" with the group by using the wildcard features of the HEAD request type to reconcile their local document knowledge with the rest of the group and can, where necessary, initiate a brief extra offer phase to distribute the newly resurrected peer's signatures.

## 3.6 User Interface

The user interface portion of the ddnfs system consists of one application that mounts the overlay file system. This file system provides all the standard file access mechanisms as per the POSIX standard[22], and operations within the overlay are handed through to a local copy of the ddnfs-controlled documents that this peers currently stores. This is completely transparent to users or their applications: new files are created using the normal creat() system call, reading and writing is handled by open(), read() and write() and so on. The file system behaviour only deviates from the POSIX standard where unsupported operations are concerned, i.e. creating symbolic links is not supported.

As documents are immutable once created, the ddnfs file system buffers newly created files until they are closed, and after an additional short settling period has elapsed without further operations on the file, the file is signed and offered to the peers and enters the normal distribution procedure as a pending document. Access attempts for documents that are not locally available but known to other peers are intercepted and elicit a document retrieval operation from the peer group.

Access to certification states and any older versions of documents is provided by a virtual directory for each document (which operates in a fashion similar to the well-known /proc file system on UNIX systems[23]) that provides read-only access to the document's signatures and various status information.

Administrative operations are handled by a separate, small stand-alone application which primarily deals with creation of administrator signatures and the management of the peerlist and policy document.

# 4  Threats and Countermeasures

## 4.1 Subverted Peers

Inadvertent information disclosure or leakage is not a major issue for a notary system, and if document contents need to be kept secret and are therefore opaque and encrypted then that does not negatively affect integrity assurances that ddnfs can provide for said document.

Assuming that a peer becomes completely subverted by an adversary, then the adversary will only gain information about what role this very computer system has played as a notary but will not be able to sabotage the global notary group's function.

No entities outside the group can impersonate a peer unless the peer's private key is compromised and the network address can be spoofed. Similarly impersonating as an administrator requires this administrator's private key.

As to providing storage to other peers, a corrupt peer can not modify other peer's data without detection, because of the self-verifying nature of the documents' plus their signature trees. It can, however, maliciously fail to respond to document retrieval requests, respond negatively or it could serve outdated (but valid) data. This can be countered by requesting a document from at least one more peers than the number of rogues to be tolerated: non-response or negative response by rogue peers will be offset by correct responses from at least one correct peer, and serving old data will be detectable by comparing the document's version in all the responses.

Offering invalid material to other peers will get the faulty peer cut off immediately: if the material is not signed or signed by the faulty peer as originator but with a broken signature, the offer will not be accepted by any other peer and the originator will be blacklisted. Injecting material overriding some other peer's documents will fail as well: overwriting existing documents is not supported. Superseding other peer's documents can be made impossible by means of the policy: the acceptable originators for a sub-namespace (e.g. the leading component of a document path name) can be limited to a legitimate originator and some administrators.

Flood-injection of valid material as a denial of service attack is possible, but a simple rate-limiting mechanism for accepting new documents by an originator suffices to limit adverse effects. Also, creating a document and its signatures and offering that is about as costly computation-wise as verifying its signatures, so the flooding attacker does not really have any advantage here.

Trying to replay previously gathered offers in a flooding attack will be detectable by rate-limiting as well: there is only a limited number of different offer messages possible for a document, so detectable repetition will set in soon.

A subverted peer can produce a little bit more obstruction by cooperating partially during the distribution process: non-communication and silently discarding offers, injecting conflicting versions of a document or propagating stripped-down offers are possible.

A small number of peers not forwarding offers (maliciously or by being out of communication) is countered by sending all offers redundantly to many peers. A document will not be replicated to all the correct peers, if and only if every correct peer handling the document chooses only rogue peers to offer to in all phases of the propagation. This is exceedingly unlikely. Also failure to participate despite being offered material will be detectable over time from analysis of the signature sets of replicated documents: if a peer remembers offering a document to a suspect peer, but never sees the suspect peer's signature on any document it receives back from the peer group, then the suspect peer's status is certainly beyond merely suspect.

A peer can also inject multiple conflicting documents with the same name and version, but this will be detected during the propagation of the documents: as soon as another peer receives an offer for a document which is in itself consistent and valid but conflicts with another document it already has stored, the culprit is known: only the rogue originator can have produced the originator's signature on these multiple conflicting documents. The peer detecting such a conflict will blacklist the originator and push the conflicting version to all the peers which store either version of the documents (which are known from the signature lists) and thus the perpetrator will be cut off from the peer group totally and rather quickly.

The most insidious and subtle obstruction would be to forward offers but strip them (partially) of their interesting content: other peers' signatures. A forwarded offer always must have at least the originator's signature and the forwarding peer's signature in it, but signatures collected in between could be stripped away. Hierarchical overlapping signatures make such stripping impractical, but it cannot be avoided completely in a minimal trust environment like this. In any case the effect of such a modification is a slowdown of propagation speeds only: any valid offer to a peer is useful. If the contacted peer is learning of a new document by a (possibly mangled) offer, then it will start propagating and collecting information about the document itself and eventually acquire any stripped signatures from other correct peers in due order. If the stripped offer does not convey new information to the contacted peer, the redundancy employed in offering documents will eventually lead to other peers offering this missed information.

### 4.2 Rogue Administrator

The situation of a rogue administrator is more problematic, depending on the nature of the policy restrictions in place.

If all peers aggressively remove superseded documents, and if there are no rules requiring multiple administrators to sign off on documents to become active, then this administrator will be able to supersede/remove all peer's data, but these changes would be detectable and traceable to the administrator because of the requirement for signatures on all replicated data.

In reality, however, a lone administrator will usually have other, more direct methods of destroying such a distributed system, so trying to guard too extensively against a person who has access to the underlying operating system anyway is rather futile.

Nevertheless ddnfs supports hierarchical administrator capabilities to facilitate operation of large installations with multiple personnel of varying trustworthiness: in such a situation the administrators can formulate a policy which limits the actions of "junior" administrators to contain potential mischief, for example by restricting their actions to changes affecting particular peers only etc. For example changes to documents of global scope like the peerlist could be set to require agreement by all senior administrators.

## 5 Implementation

Our prototype implementation was written in Perl, mainly for the usual reasons: Perl's suitability for rapid development, the extreme conciseness of the written code and an abundance of existing Perl packages for interfacing with just about any established technology.

The first two aspects, however, are also causing the testbed implementation to be somewhat too slow for wide-spread deployment: Perl has a considerable memory and CPU footprint, and performing large amounts of cryptographic operations doesn't exactly help. Nevertheless Perl proved to be an exceptionally useful development platform because very mature packages could be used for covering the core cryptographic and file system aspects of the system.

The ddnfs application makes heavy use of the excellent FUSE infrastructure[24], a mechanism that lets one implement a file system as an application in user space without having to modify the operating system kernel. FUSE implementations are available for most POSIX-compliant operating systems, and the FUSE-Perl interface worked very well and provided reasonable performance.

ddnfs is a multi-threaded application, with one thread marshalling communications with other peers and any number of threads performing the overlay file system operations. The operations within the overlay file system are translated into accesses to local copies of the documents that the particular peer stores where possible (e.g. reading an existing document's data). Whenever interactions with other peers are required (e.g. retrieval of previously unknown documents or distribution of newly created material) the marshalling thread is contacted which in turn creates and manages any number of communicant threads that handle one peer-to-peer connection.

The cryptographic components of ddnfs were implemented using SSL/TLS, specifically the OpenSSL implementation and its Perl interface, because of its maturity and the consequential ease of establishing safely authenticated and encrypted communication sessions with SSL. The SSL trust infrastructure is not used but instead individual peers' public keys are included in the peerlist in the form of selfsigned certificates. At this time the low-level communications are performed over SSL-encrypted TCP connections, but this could be extended quite easily to use any type of authenticated and encrypted channel (e.g. RESTful web service or even email with OpenPGP).

## 6 Conclusion

In this paper we describe a novel approach for using independent parties as automated online notaries for providing document integrity certifications that are robust and tamper-resistant against malicious failures or subversion of a limited number of peers. This decentralized peer-to-peer setup requires extremely limited trust assumptions and therefore performs much more flexibly than existing online certification infrastructures. At the same time ddnfs also presents a very low entry barrier to potential adopters due to its simple user interface that makes the certified and replicated data accessible in the form of an overlay file system and which makes using ddnfs transparent to users and their applications.

A prototype implementation was developed and used to demonstrate the effectiveness and practical boundaries of our notary system, which does not require special-purpose cryptographic hardware or trusted networks for its operation, supports decentralized control by multiple independent administrators, and offers benefits to any application which needs tamper-resistant, replicated storage with read-oftenwrite-seldom semantics. Based on these characteristics, we believe that ddnfs demonstrates that a successful balance between paranoid mistrust and practical efficiency is possible for a data integrity certification system and that its approach to trust and integrity assurance is well attuned to the necessities of today's decentralized online environments.